\documentclass[pdflatex,sn-mathphys-num,iicol]{sn-jnl}

\usepackage{graphicx}%
\usepackage{multirow}%
\usepackage{amsmath,amssymb,amsfonts}%
\usepackage{amsthm}%
\usepackage{mathrsfs}%
\usepackage[title]{appendix}%
\usepackage{xcolor}%
\usepackage{textcomp}%
\usepackage{manyfoot}%
\usepackage{booktabs}%
\usepackage{algorithm}%
\usepackage{algorithmicx}%
\usepackage{algpseudocode}%
\usepackage{listings}%
\usepackage{cuted}%

\theoremstyle{thmstyleone}%
\theoremstyle{thmstyletwo}%

\theoremstyle{thmstylethree}%

\begin{document}

\title{Energy-flux noise induced by imperfections of the Gaussian beam in the Li-Baker HFGW detector}


\author[1]{Lin-Lin Yang}\email{d202380057@hust.edu.cn}
\author[2]{Hao Yu}\email{yuhaocd@cqu.edu.cn}
\author*[2]{Pan-Pan Wang}\email{ppwang@cqu.edu.cn}
\author[2]{Jin Li}\email{cqujinli1983@cqu.edu.cn}
\author[3]{Cheng-Gang Shao}\email{cgshao@yzu.edu.cn}

\affil[1]{National Gravitation Laboratory, MOE Key Laboratory of Fundamental Physical Quantities Measurement, and School of Physics, Huazhong University of Science and Technology, Wuhan 430074, People's Republic of China}

\affil[2]{College of Physics, Chongqing University, Chongqing 401331, China}

\affil[3]{School of Physics and Optoelectronic, Yangtze University, Jingzhou 434023, China}


\abstract{\unboldmath
This paper investigates the energy-flux noise induced by imperfections of
the Gaussian beam (GB) in the Li-Baker high-frequency gravitational wave (HFGW) detector.
Modeling these imperfections with higher-order Laguerre–Gaussian (LG) beams,
we analytically derive the resulting noise-to-signal ratio for both co- and
cross-polarized $\mathrm{LG}_{nl}$ modes. It is found that higher-order
co-polarized LG beams generate no tangential energy-flux noise at the optimal
detection plane ($\phi=\pi/2$ or $3\pi/2$), whereas higher-order cross-polarized
LG beams introduce non-negligible noise. Specifically, for an individual higher-order
cross-polarized $\mathrm{LG}_{nl}$ mode under our representative parameters,
maintaining the noise-to-signal ratio below unity typically requires the power
fraction to satisfy $\sigma_{nl}<10^{-46}$, with the $\mathrm{LG}_{01}$ mode imposing
the most stringent constraint of $\sigma_{01}<6.0\times10^{-49}$. {Monte Carlo simulations
suggest that the total power fraction $\sigma_{\mathrm{total}}$  of multiple higher-order cross-polarized $\mathrm{LG}_{nl}$ modes
needs to remain below $10^{-49}$}. Consequently, we propose employing a polarization filter to retain only the co-polarized component of the GB adopted in the practical experiment, thereby suppressing the noise while preserving signal integrity. These results provide theoretical guidance for noise control in the Li-Baker HFGW detector.}




\maketitle

\section{Introduction}\label{sec1}
In 1916, Einstein formulated general relativity and predicted the existence of
gravitational waves (GWs). In 2015, the LIGO Scientific Collaboration
and the Virgo Collaboration directly detected a GW from a binary black hole merger
for the first time~\cite{Abbott2016GW150914}. This detection not only confirmed
a key prediction of general relativity but also ushered in the era of GW astronomy.
Over the past decade, a growing body of work has shown that GWs encode rich cosmological
and astrophysical information, including mergers of binary black holes~\cite{Abbott2016GW150914,GW170104},
binary neutron stars~\cite{Abbott2017GW170817},
cosmic inflation~\cite{Massimo1999PRDQIM,gravitational_waves_from_inflation_2016},
and extra dimensions~\cite{Andriot2017JCAPED,Yu_2019},
among others. The information carried by GWs complements electromagnetic observations and thus opens new avenues for investigating fundamental physics and cosmic evolution. Developing sensitive and reliable methods for detecting GWs is therefore an important objective in modern astrophysics and gravitational physics.

In theory, GWs span a vast frequency range and are conventionally divided into several
bands, each associated with distinct sources and detection
methods~\cite{Moore_2015,Caprini_2018}.
Extremely low frequencies ($10^{-18}$--$10^{-16}\,\mathrm{Hz}$) are probed
through the polarization of the cosmic microwave background and
carry information about inflation~\cite{Massimo1999PRDQIM,gravitational_waves_from_inflation_2016,annurev:/content/journals/10.1146/annurev-astro-081915-023433}.
Nanohertz waves ($10^{-9}$--$10^{-7}\,\mathrm{Hz}$) are accessible
to pulsar timing arrays~\cite{Burke_Spolaor_2019},
millihertz waves ($10^{-4}$--$1\,\mathrm{Hz}$) to
space-based interferometers~\cite{LISA}
, and waves from about $1\,\mathrm{Hz}$
to $10^{4}\,\mathrm{Hz}$ to ground-based detectors such as LIGO,
Virgo, and KAGRA, among others~\cite{LISA,LIGO1,LIGO2,Virgo,KAGRA,ET1,ET2,taiji,TianQin}.
Still higher frequencies, from the kilohertz band into the gigahertz range,
remain largely unexplored and are potential targets of
proposed laboratory-scale detectors~\cite{Aggarwal_2025}.

High-frequency gravitational waves (HFGWs) are a natural prediction of various
early-universe processes, including first-order phase
transitions~\cite{Caprini_2016},
preheating after inflation~\cite{Easther_2006},
and the evaporation of light primordial black holes~\cite{PhysRevD.84.024028}.
Their detection would provide access to physics at energy scales beyond the reach
of electromagnetic observations. To cover various frequency bands of HFGWs,
a variety of experimental schemes have been proposed or are currently under
development~\cite{Aggarwal_2025}.
The Fermilab Holometer, a pair of Michelson interferometers
with arms of about 40 m, has set limits on a stochastic background
between 1 and 13 MHz~\cite{PhysRevD.95.063002}.
LIGO, Virgo, KAGRA and GEO600 retain a comb of optical sensitivity at
integer multiples of the free spectral range of their arm cavities,
extending into the MHz and even the GHz
range~\cite{schnabel2024opticalsensitivitiescurrentgravitational}.
Between a few tens and a few hundred kilohertz, optically levitated dielectric microdiscs or
microspheres can serve as frequency-tunable, high-quality-factor resonant sensors;
the Levitated Sensor Detector is being built
along this line~\cite{PhysRevLett.110.071105,PhysRevLett.128.111101}.
Near the megahertz, cryogenic quartz bulk-acoustic-wave resonators
use high-order acoustic modes as strain antennas to detect GWs,
read out with SQUIDs~\cite{PhysRevD.90.102005}.
Above roughly a gigahertz, a more promising route is the conversion of GWs
into photons in a strong magnetic field, i.e., the inverse Gertsenshtein
effect ~\cite{Gertsenshtein1962JETP,Boccaletti1970NC,Aggarwal_2025,PhysRevD.37.1237}.
The process is analogous to axion–photon Primakoff conversion, so many axion
experiments can be turned to gravitational-wave searches
by reinterpreting existing data or by modest changes in geometry~\cite{PhysRevLett.129.041101,PhysRevD.105.116011,PhysRevD.111.035031}.
Because all these GW detectors are ultra-high-precision measuring instruments,
rigorous modeling and analysis of all potential noise sources are essential.
In this paper, we focus on a HFGW detector based on the inverse Gertsenshtein effect,
often referred to as the Li-Baker (HFGW) detector~\cite{Li2003PRD,Li1999CPL,Li2000PRD,Li2002IJMPD,LiBaker2007IJMPB,Li2008EPJC,Stephenson2009AIP}. This detection scheme has been further developed and refined over the past decade~\cite{Li2009PRD,Li2016NPB,Li2020EPJC,Li2023PRD,Zheng2018PRD,Wu2023CPC,Wen2014PRD,TongZhang2009PRD}.

\begin{figure}[t]
    \centering
    \includegraphics[width=0.5\textwidth]{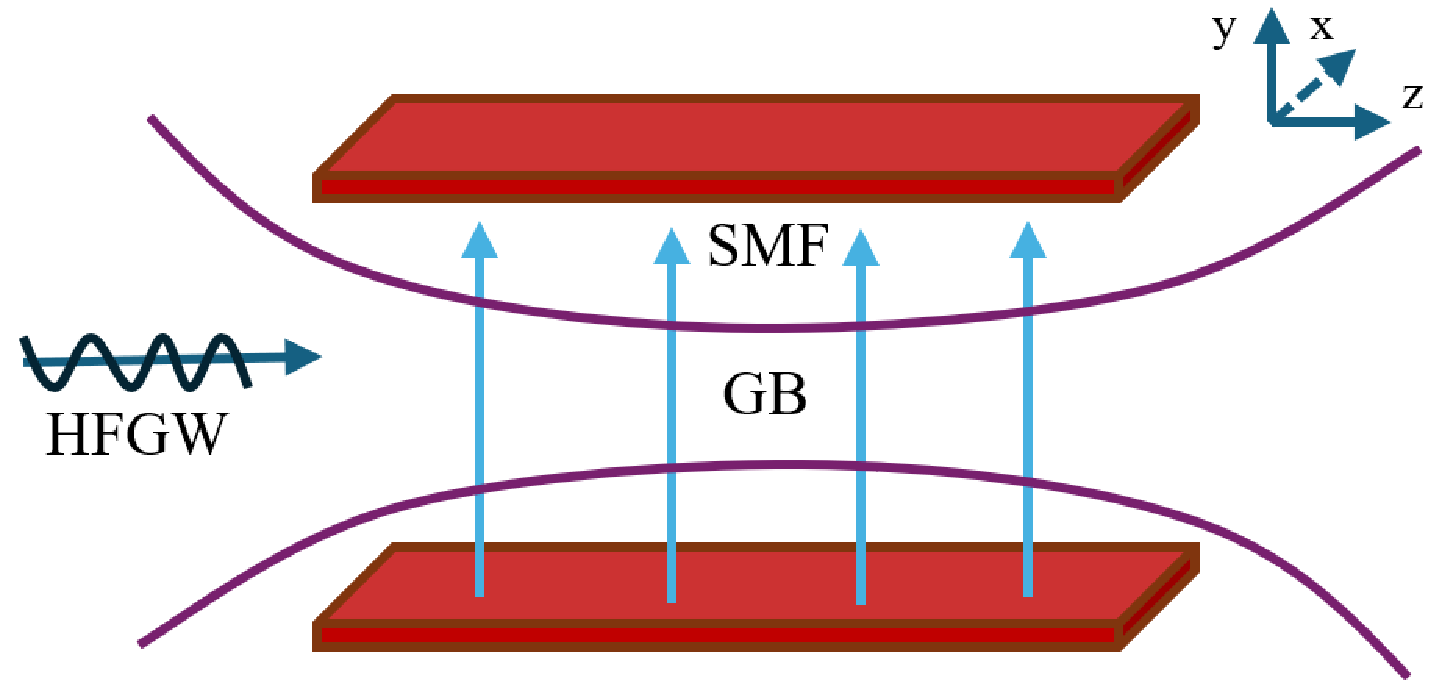}
    \caption{Schematic diagram of the Li-Baker detector. Here, SMF abbreviates the (background) static magnetic field, and the purple curve represents the profile of the GB.}    \label{fig:f1}
\end{figure}

The electromagnetic wave induced by the GW via the inverse Gertsenshtein
effect is directly proportional to the amplitude of the GW. Consequently,
the energy flux of the induced electromagnetic wave is proportional to
the square of the GW amplitude. Since the GW strain is extremely small,
typically of order $10^{-30}/\mathrm{\sqrt{Hz}}$,
its square is vanishingly small, making the direct detection of
GW signals extremely challenging via the inverse Gertsenshtein effect. In the Li-Baker detector, a Gaussian beam (GB) is introduced to propagate along the \(z\)-direction through a static strong magnetic field, and the magnetic field can be oriented along arbitrary directions to serve different detection purposes (see Fig.~\ref{fig:f1}). Under the synchro-resonance condition, in which the GB and the target HFGW have the same frequency, the HFGW interacting with the static magnetic field induces a first-order perturbative electromagnetic field. Coherent interference of this electromagnetic field with the GB produces a transverse first-order energy flux proportional to the GW amplitude. The scheme proposes to search for that energy flux on the longitudinal symmetry plane of the GB, where the transverse background photon flux is predicted to be strongly suppressed~\cite{Li2003PRD}.

In the Li–Baker HFGW scheme, the noise budget has been examined repeatedly in the literature. The background photon flux is commonly taken to be the dominant contribution. Diffraction and shot noise have been simulated to lie below the signal, thermal noise can be reduced by cooling the microwave receiver, and quantum noise is not regarded as limiting. These conclusions, however, are based on ideal conditions, including a perfectly static magnetic field and a pure (perfect) GB, which a real-world detector cannot provide. Even the laser of LIGO, the most sensitive GW detector now in operation, retains residual power fluctuations, and the purity of its fundamental Gaussian mode is only 99.5\%~\cite{Kwee:12}. Obviously, imperfections in the static magnetic field or the GB may enhance the noise-to-signal ratio in the Li–Baker HFGW scheme. Motivated by this concern, we study the noise induced by imperfections of the GB. In this paper, these imperfections are modeled as higher-order Laguerre–Gaussian (LG) beams. We derive the induced energy-flux noise analytically and obtain the higher-order-mode power fractions compatible with the detection of HFGWs.

The paper is organized as follows. In Sec.\ref{sec2}, we briefly review the operating principle of the Li-Baker detector. In Sec.\ref{sec3}, we develop the energy-flux noise model with higher-order LG beams. In Sec.\ref{sec4}, we present the numerical analyses, determine the allowable power limits for LG beams, and discuss noise suppression methods. Finally, conclusions are summarized in Sec.\ref{sec5}.

\section{Operating principles of the Li-Baker detector}\label{sec2}

Evaluating the noise-to-signal ratio is a crucial step in assessing the feasibility of GW detectors. Because the GW signal is extremely weak, any potential background noise must be rigorously compared with the expected signal level to determine whether it would overwhelm the signal and prevent detection. This study focuses on the energy-flux noise arising from imperfections of the Gaussian beam (GB) in the Li-Baker detector and analyzes how this noise affects the noise-to-signal ratio. To this end, we first review the operating principles of the Li-Baker detector based on Ref.~\cite{Li2003PRD}.

Since weak GWs can be treated as a small perturbation on the background spacetime, in the transverse-traceless gauge, the metric tensor of a monochromatic circular polarized plane GW can be expressed as
\begin{align}{d} s^2&=-c^2d t^2+d z^2+\left[1+h_+(z,t)\right]d x^2\notag\\
&+\left[1-h_+(z,t)\right]d y^2+2h_\times(z,t)dxdy.\label{metric}
\end{align}
where $h_+(z,t)$ and $h_\times(z,t)$ represent the ``+'' and ``$\times$''polarization of the GW propagating along the $z$ direction, respectively. In this study, we restrict our attention to this simplest configuration of GWs. The methodology can be straightforwardly generalized to other possible polarization states. These two polarization states correspond to tensor-mode gravitons, and can be written as
\begin{align}
    h_+(z,t)=A_+e^{-i\omega_gt+ik_gz},
\\
    h_\times(z,t)=A_\times e^{-i\omega_gt+ik_gz}.
\end{align}
Here, $A_+$ and $A_\times$ represent the amplitudes of $h_+(z,t)$ and $h_\times(z,t)$, respectively,
with $\omega_g$ and $k_g$ being the angular frequency and wavenumber of the GW.

To study the electromagnetic response to the GW in the Li-Baker detector, we need to employ the Maxwell equations in a curved spacetime background, which take the following form:
\begin{align}
\frac{1}{\sqrt{-g}}\partial_\nu
\left(\sqrt{-g}\,g^{\mu\alpha}g^{\nu\beta}F_{\alpha\beta}\right)&=\mu_0 J^\mu,\\
\partial_{[\alpha}F_{\beta\gamma]}&=0 .
\end{align}
Here, $F_{\alpha\beta}$ represents electromagnetic field tensor and
$J^\mu$ denotes four-dimensional electric current density. In the context of the
Li-Baker detector, the electromagnetic response is assumed to take place in a vacuum,
and thus $J^\mu=0$~\cite{Grishchuk:1975tg}. The inverse metric tensor $g^{\mu\alpha}$ and the corresponding determinant $g$ are given by Eq.~\eqref{metric}. Moreover, $F_{\alpha\beta}$ comprises three parts: the background static magnetic field $\hat{F}^{(0)}_{\alpha\beta}$, the
background electromagnetic field $\tilde{F}^{(0)}_{\alpha\beta}$ (i.e., the background GB), and the induced first-order perturbation $\tilde{F}^{(1)}_{\alpha\beta}$ (corresponding to the target electromagnetic signals to be detected) arising from the background fields in the presence of the GW. Mathematically, this can be expressed as
\begin{align}
    F_{\alpha\beta}=\hat{F}^{(0)}_{\alpha\beta}+\tilde{F}^{(0)}_{\alpha\beta}+\tilde{F}^{(1)}_{\alpha\beta}.
\end{align}
Here, we adopt the same notation as in Ref.~\cite{Li2003PRD}. Specifically, ``0'' denotes the background electromagnetic fields, ``1'' denotes the first-order perturbation, a hat ``$\,\,\hat{ }\,\,$'' denotes the static electromagnetic field, and a tilde ``$\,\,\tilde{ }\,\,$'' represents the time-dependent electromagnetic field.

In the Li-Baker detector, the first-order perturbation $\tilde{F}^{(1)}_{\alpha\beta}$ induced by GWs of different polarization modes exhibits distinct characteristics. According to Ref.~\cite{Li2003PRD}, only the electromagnetic signal induced by the ``$\times$''-mode GW can be detected. In contrast, the electromagnetic signal induced by the ``$+$''-mode GW is overwhelmed by the energy flux of the background electromagnetic field, rendering it undetectable. Considering the ``$\times$''-mode GW and neglecting the perturbative effect arising from the direct interaction between the GW and the GB, the Maxwell equations yield
\begin{align}
    \Box \tilde{E}^{(1)}_y&=\tilde{E}^{(1)}_{y,zz}-\frac{1}{c^2}\tilde{E}^{(1)}_{y,tt}\notag\\
    &=-iA_\times \hat{B}^{(0)}_yk_g^2c\,e^{i(k_gz-\omega_g t)},\label{Ey}
\\
    \Box \tilde{B}^{(1)}_x&=\tilde{B}^{(1)}_{x,zz}-\frac{1}{c^2}\tilde{B}^{(1)}_{x,tt}\notag\\
    &=iA_\times \hat{B}^{(0)}_yk_g^2e^{i(k_gz-\omega_g t)},\label{Bx}
\\
   \tilde{E}^{(1)}_x & =\tilde{B}^{(1)}_y= \tilde{E}^{(1)}_z= \tilde{B}^{(1)}_z=0.
\end{align}
The electric field component $\tilde{E}^{(1)}_y$ and magnetic field component $\tilde{B}^{(1)}_x$ constitute the non-vanishing components of the induced first-order perturbation $\tilde{F}^{(1)}_{\alpha\beta}$. Note that $\hat{B}^{(0)}_y$ represents the strength of the background static magnetic field along the $y$-axis. In this case, the background static magnetic field can be assumed to be oriented solely along the $y$-axis ($\hat{B}^{(0)}=\hat{B}^{(0)}_y$), with an effective range of $-\ell/2\leq z\leq \ell/2$ along the $z$-axis. The general solutions of Eqs.~\eqref{Ey} and~\eqref{Bx} are given by
\begin{align}
\tilde{E}_y^{(1)} &= - \frac{1}{2} A_\times \hat{B}_y^{(0)} k_g c (z + \ell/2) \exp[i(k_g z - \omega_g t)] \nonumber\\
& + i b_y \exp[i(k_g z - \omega_g t)] + i d_y \exp[i(k_g z + \omega_g t)], \\
\tilde{B}_x^{(1)} &= \frac{1}{2} A_\times \hat{B}_y^{(0)} k_g (z + \ell/2) \exp[i(k_g z - \omega_g t)] \nonumber\\
& + i b_x \exp[i(k_g z - \omega_g t)] + i d_x \exp[i(k_g z + \omega_g t)].
\end{align}
The exact expressions of the four constants are specified by the physical conditions and boundary constraints, and they must satisfy
\begin{align}
b_y + c\, b_x = -\frac{1}{2} A_\times \hat{B}_y^{(0)} c, \qquad d_y - c\, d_x = 0.
\end{align}

In curved spacetime, only local measurements along an observer's worldline possess definitive physical meaning. These observable quantities correspond to the projections of tensor physical quantities onto the observer's tetrads $\tau^\mu_{(a)}$, which comprise a timelike vector parallel to the observer's four-velocity and three mutually orthogonal spacelike vectors. Here, the index in parentheses acts as a label for the vectors, whereas the remaining index specifies the coordinate components of the tetrads. When only the ``$\times$''-mode GW is present, the observer's tetrad $\tau^\mu_{(a)}$ in the Li-Baker detector can be expressed as
\begin{equation}
\tau^\mu_{(a)}=
\begin{pmatrix}
1 & 0 & 0 & 0 \\
0 & 1 & 0 & 0 \\
0 & -h_\times & 1 & 0 \\
0 & 0 & 0 & 1
\end{pmatrix}.
\end{equation}
Therefore, the quantities $F_{(ab)}$ measured by the observer are given by
\begin{align}
    F_{(ab)}=F_{\mu\nu}\tau^\mu_{(a)}\tau^\nu_{(b)}.
\end{align}
Neglecting high-order infinitely small quantities, one can find that for the ``$\times$''-mode GW, the components $\tilde{E}_{(y)}^{(1)}$ and $\tilde{B}_{(x)}^{(1)}$ of $F_{(ab)}$ measured by the observer are given by
\begin{align}
\tilde{E}_{(y)}^{(1)} &=\tilde{E}_y^{(1)}\nonumber\\
&= - \frac{1}{2} A_\times \hat{B}_y^{(0)} k_g c (z + \ell/2) \exp[i(k_g z - \omega_g t)] \nonumber\\
& + i b_y \exp[i(k_g z - \omega_g t)] + i d_y \exp[i(k_g z + \omega_g t)],
\\
\tilde{B}_{(x)}^{(1)} &=\tilde{B}_x^{(1)}+h_{\times}\hat {B}_y^{(0)}\nonumber\\
&= \frac{1}{2} A_\times \hat{B}_y^{(0)} k_g (z + \ell/2) \exp[i(k_g z - \omega_g t)] \nonumber\\
& + i b_x \exp[i(k_g z - \omega_g t)] + i d_x \exp[i(k_g z + \omega_g t)]\nonumber\\
&+ i A_{\times} \hat{B}_y^{(0)} \exp[i(k_g z - \omega_g t)].
\end{align}
The above equations represent the general solutions to the Maxwell equations in the context of the Li-Baker detector. Taking into account the specific boundary conditions ($z \leq -\ell/2, \hat{B}^{(0)} = 0$; $-\ell/2 \leq z \leq \ell/2, \hat{B}^{(0)} = \hat{B}_y^{(0)}$; $\ell/2 \leq z \leq \ell_0, \hat{B}^{(0)} = 0$ with $\ell_0$ being characteristic size for plane-wave behavior of second-order electromagnetic fluxes), one can obtain the particular solutions corresponding to these boundary conditions. Taking the real part of the purely perturbative fields, the particular solutions can be expressed as
\begin{align}
\tilde{E}_{(y)}^{(1)} &= -\frac{1}{2} A_{\times} \hat{B}_{y}^{(0)} k_g c\left(z+\ell/2\right)\cos\left(k_g z - \omega_g t\right) \notag\\
&-\frac{1}{2} A_{\times} \hat{B}_{y}^{(0)} c \sin\left(k_g z\right)\cos\left(\omega_g t\right),
\\
\tilde{B}_{(x)}^{(1)} &= \frac{1}{2} A_{\times} \hat{B}_{y}^{(0)} k_g\left(z+\ell/2\right)\cos\left(k_g z - \omega_g t\right) \notag\\
& -\frac{1}{2} A_{\times} \hat{B}_{y}^{(0)} \sin\left(k_g z\right)\cos\left(\omega_g t\right).
\end{align}

Although the electromagnetic field has physical significance, it is not a direct observable. In the Li-Baker detector, the observables are instead the energy fluxes arising from the interaction between the induced electromagnetic field and the background GB. The energy-momentum tensor of the electromagnetic field $F_{\mu \nu}$ in the gravitational field can generally be expressed as
\begin{align}
    T^{\mu\nu} \equiv \frac{1}{\mu_0} \left( -F^\mu{}_\alpha F^{\nu\alpha} + \frac{1}{4} g^{\mu\nu} F_{\alpha\beta} F^{\alpha\beta} \right),
\end{align}
which can be decomposed into
\begin{align}
   T^{\mu\nu} = \overset{(0)}{T^{\mu\nu}} + \overset{(1)}{T^{\mu\nu}} + \overset{(2)}{T^{\mu\nu}}.
\end{align}
Here, \(\overset{(0)}{T^{\mu\nu}}\) corresponds to the background electromagnetic field, while \(\overset{(1)}{T^{\mu\nu}}\) and \(\overset{(2)}{T^{\mu\nu}}\) represent the first- and second-order perturbations to \(\overset{(0)}{T^{\mu\nu}}\) induced by the GW. In cylindrical polar coordinates, the first-order radial, tangential, and axial perturbative power flux densities (energy flux) satisfy, respectively,
\begin{align}
   \overset{(1)}{S^r} = c \overset{(1)}{T^{01}},\,\,\,\,\,\overset{(1)}{S^\phi} = c \overset{(1)}{T^{02}},\,\,\,\,\,\,\overset{(1)}{S^z} = c \overset{(1)}{T^{03}}.
\end{align}

For the Li-Baker detector, we focus solely on the first-order tangential average power flux density $\overset{(1)}{\langle S^\phi \rangle}_{\omega_e = \omega_g}$ in cylindrical polar coordinates under the synchro-resonance condition $\omega_e = \omega_g$. For the ``$\times$''-mode GW,
the induced tangential averaged power flux density at coordinates  ($r$,$\phi$) in the plane $z=0$ is finally expressed as~\cite{Li2003PRD}
\begin{align}\label{S_ideal}
    \overset{(1)}{\langle S_\times^\phi \rangle}_{\omega_e = \omega_g}
\equiv
\frac{A_{\times} \hat{B}_{y}^{(0)} \psi_0 \,\ell\, r}{4 \mu_0 W_0^2}
\exp\left( -\frac{r^2}{W_0^2} \right)
\sin^2 \phi ,
\end{align}
where $W_0$ is the spot radius of the GB at $z=0$ and
$\psi_0$ is the amplitude of the GB at $z=0$. With the power flux density, one can perform an area integration to obtain the total perturbative power and the photon flux. Combining these with the noise level and integration time, the noise-to-signal ratio can then be estimated to determine whether the GW is truly detectable.

\section{Laguerre--Gauss beams and energy flux noise}\label{sec3}
In Sec.~\ref{sec2}, we reviewed the operating principles of the Li-Baker detector in the context of an ideal GB. In reality, it is practically impossible for any optical beam to achieve 100\% purity. LIGO, one of the most advanced GW detectors, employs a laser beam in which the fundamental Gaussian mode accounts for approximately 99.5\% of the total power~\cite{Kwee:12}. A similar situation applies to the Li-Baker detector,
where the energy fraction of the fundamental Gaussian mode cannot reach 100\% in practice either. If imperfect GBs are present in the Li-Baker detector, their interaction with the fundamental (ideal) GB will also generate an energy flux, acting as an additional noise source. If this energy flux exceeds the one induced by the GW, the noise-to-signal ratio of the system may be enhanced, depending on whether the noise caused by the non-standard GBs is dominant compared to other noise sources. In the following, we study the impact of imperfect GBs on the noise-to-signal ratio of the Li-Baker detector by employing Laguerre-Gaussian (LG) beams to model these non-ideal profiles. Meanwhile, other noise sources are temporarily disregarded in this analysis. Note that, for fixed $r$ and $z$, the induced tangential averaged power flux density generally attains its maximum at $\phi=\pi/2$ or $3\pi/2$ [see Eq.~\eqref{S_ideal}]. The optimal detection planes are therefore expected to lie at these azimuthal angles. In what follows, we restrict our attention to the noise-to-signal ratio on the planes corresponding to $\phi=\pi/2$ or $3\pi/2$.

\subsection{Laguerre-Gaussian beams}\label{subsec31}
In the cylindrical coordinate system, the normalized LG beams can be expressed as~\cite{PhysRevA.45.8185}
\begin{equation}
\mathrm{LG}_{nl}(r,\phi,z)=A_{nl}(r,z)e^{i\Phi_{nl}(r,\phi,z)},
\label{eq:Enl}
\end{equation}
where
\begin{align}
A_{nl}(r,z)
&=
\sqrt{\frac{2n!P_{nl}}{\pi(n+|l|)!}}\,
\frac{1}{W(z)}
\left(\frac{\sqrt{2}r}{W(z)}\right)^{|l|}\notag\\
&\times L_n^{|l|}\!\left(\frac{2r^2}{W^2(z)}\right)
\exp\left[-\frac{r^2}{W^2(z)}\right],
\label{eq:Anl}
\end{align}
and
\begin{align}
\Phi_{nl}(r,\phi,z)
=
kz-\mathrm{gp}(z)+\frac{kr^2}{2R(z)}-l\phi+\theta_{nl}.
\label{eq:Phinl}
\end{align}
Here, \(n\) and \(l\) represent the radial index and the azimuthal index, respectively.
\(L_n^{|l|}\) is the general Laguerre polynomial.
\(P_{nl}\) denotes the power carried by the \((n,l)\) mode and \(\theta_{nl}\)
is its arbitrary phase.
$\mathrm{gp}(z)$ represents Gouy phase and $R(z)$ denotes the radius of curvature. The beam waist radius $W(z)$ is given by
\begin{align}
    W(z)= W_0\sqrt{1+\left(z\lambda_e/\pi W_0^2\right)^2},
\end{align}
where $\lambda_e=2\pi c/\omega_e$ is the beam wavelength and $W_0$ is the beam waist radius at $z=0$.

When $(n,l)=(0,0)$, Eq.~\eqref{eq:Enl} reduces to the fundamental GB expressed as
\begin{align}\label{GB}
    \mathrm{LG}_{00}=\sqrt{\frac{2P_{00}}{\pi}}\,
\frac{1}{W(z)}\exp\left[-\frac{r^2}{W^2(z)}\right]e^{i\Phi_{00}}.
\end{align}
Therefore, all modes with $(n,l) \neq (0,0)$ can be treated as imperfect GBs. Hereafter, we refer to all LG beams other than $\mathrm{LG}_{00}$ as higher-order LG beams. Note that with the aid of Eq.~\eqref{GB}, the amplitude $\psi_0$ of the GB at $z=0$ in Eq.~\eqref{S_ideal} is consequently determined by
\begin{align}\label{psi0}
    \psi_0=\sqrt{\frac{2P_{00}}{\pi W_0^2}}.
\end{align}

We define the power fraction of higher-order LG beams as
\begin{equation}
\sigma_{nl}=\frac{P_{nl}}{P_{00}}\ll 1 .
\label{eq:sigma}
\end{equation}
The field amplitude of the $(n,l)$ mode is proportional to \(\sqrt{\sigma_{nl}}\).
It should be noted that the polarization state of the higher-order LG beam is not specified in Eq.(\ref{eq:Enl}). In this work, we focus on the standard $x$- and $y$-polarized states and analyze their energy-flux noise. Nevertheless, more general polarization configurations, such as circularly or cylindrically polarized LG beams, can be investigated along the same lines.

\subsection{Co-polarized LG beam}\label{subsec32}
We first consider higher-order $x$-polarized LG beams, also referred to as co-polarized LG beams. In this case, the non-ideal electric field component of the background electromagnetic field $\tilde{F}^{(0)}_{\alpha\beta}$ can be expressed as
\begin{align}
    \tilde{\mathbf{E}}^{(0)}_{\mathrm{ni}}=\mathbf{\hat{x}}\left[\tilde{E}^{(0)}_{x,\mathrm{ideal}}+\sum_{(n,l)\ne(0,0)}\mathrm{LG}_{nl}\right],
\end{align}
where $\mathbf{\hat{x}}$ represents $x$-polarization.
$\tilde{E}^{(0)}_{x,\mathrm{ideal}}e^{-i\omega_e t}$ is the only nonzero component of $\tilde{F}^{(0)}_{\alpha\beta}$ in the presence of only the ideal GB. Therefore, the corresponding magnetic field component is given by
\begin{align}\label{nablaE}
    \tilde{\mathbf{B}}^{(0)}_{\mathrm{ni}}=-\frac{i}{\omega_e}\nabla \times \tilde{\mathbf{E}}^{(0)}_{\mathrm{ni}}.
\end{align}
In this case, the components of the time-averaged energy flux in the $x$- and $y$-directions resulting from higher-order $x$-polarized LG beams (i.e., the noise caused by the non-standard GBs) vanish under the synchro-resonance condition $\omega_e = \omega_g$. In the $\phi$-direction, the time-averaged energy flux depends on $\phi$ and can be written as
\begin{align}\label{Sxx}
    \left\langle S_{\phi,x}^{xx} \right\rangle&=-\frac{1}{2\mu_0} \mathrm{Im}\left[\tilde{E}^{(0)}_{r,\mathrm{ni}}\tilde{B}^{(0)}_{z,\mathrm{ni}}{}^{*} \right]\notag\\
    &=-\frac{\cos\phi}{2\mu_0}\mathrm{Im}\left[\tilde{E}^{(0)}_{x,\mathrm{ni}}\tilde{B}^{(0)}_{z,\mathrm{ni}}{}^{*} \right].
\end{align}
Here, $\tilde{E}^{(0)}_{r,\mathrm{ni}}=\cos\phi\tilde{E}^{(0)}_{x,\mathrm{ni}}$ is the radial component of $\tilde{\mathbf{E}}^{(0)}_{\mathrm{ni}}$,
and $\tilde{B}^{(0)}_{z,\mathrm{ni}} $ denotes the $z$-component of $\tilde{\mathbf{B}}^{(0)}_{\mathrm{ni}}$.
One can find that when $\phi = \pi/2$ or $3\pi/2$, the energy flux is also consistently zero. In other words, it does not contribute to the noise in the Li-Baker detector along these directions. Therefore, higher-order co-polarized LG beams are not further considered in the numerical simulations.

\subsection{Cross-polarized LG beam}
For higher-order $y$-polarized (cross-polarized) LG beams, we have
\begin{align}
    \tilde{\mathbf{E}}^{(0)}_{\mathrm{ni}}=\mathbf{\hat{x}}\,\tilde{E}^{(0)}_{x,\mathrm{ideal}}
    +\mathbf{\hat{y}}\sum_{(n,l)\ne(0,0)}\mathrm{LG}_{nl},
\end{align}
where $\mathbf{\hat{y}}$ represents $y$-polarization. Similarly, the magnetic field component satisfies
\begin{align}
    -i\omega_e\tilde{\mathbf{B}}^{(0)}_{\mathrm{ni}}&=\mathbf{\hat{x}}\partial_z\left(\sum\mathrm{LG}_{nl}\right)+
    \mathbf{\hat{y}}\partial_z\tilde{E}^{(0)}_{x,\mathrm{ideal}}\notag\\
    &+\mathbf{\hat{z}}\left[\partial_x\left(\sum\mathrm{LG}_{nl}\right)-\partial_y\tilde{E}^{(0)}_{x,\mathrm{ideal}}\right].
\end{align}
Therefore, under the synchro-resonance condition $\omega_e = \omega_g$, the components of the time-averaged energy flux in the
$x$- and $y$-directions are expressed as follows, respectively:
\begin{align}
    \left\langle S_{x,(n,l)}^{xy} \right\rangle=\frac{1}{2\mu_0\omega_e}
    \mathrm{Im}\left[\mathrm{LG}_{nl}\partial_y\tilde{E}^{(0)}_{x,\mathrm{ideal}}{}^*\right]
\end{align}
and
\begin{align}
    \left\langle S_{y,(n,l)}^{xy} \right\rangle=\frac{1}{2\mu_0\omega_e}
    \mathrm{Im}\left[\tilde{E}^{(0)}_{x,\mathrm{ideal}}\partial_x\mathrm{LG}_{nl}{}^*\right].
\end{align}
Furthermore, the time-averaged energy flux in the $\phi$-direction is given by
\begin{align}\label{Sxynl}
    &\left\langle S_{\phi,(n,l)}^{xy} \right\rangle=-\left\langle S_{x,(n,l)}^{xy} \right\rangle\sin\phi+\left\langle S_{y,(n,l)}^{xy} \right\rangle\cos\phi\nonumber\\
    &=-\frac{A_{00}A_{nl}}{2\mu_0\omega_e}\Bigg\{
\left[
\frac{k\,r}{R(z)}\cos2\phi
+\frac{l}{2r}\sin2\phi
\right]\cos\Delta_{nl}\nonumber\\
&+\left(
\frac{\partial_r A_{00}\sin^2\phi}{A_{00}}
+\frac{\partial_r A_{nl}\cos^2\phi}{A_{nl}}
\right)\sin\Delta_{nl}\Bigg\}.
\end{align}
where $\Delta_{nl}=\Phi_{nl}-\Phi_{00}$.
Unlike Eq.~\eqref{Sxx}, Eq.~\eqref{Sxynl} lacks the overall \(\cos\phi\) factor, which means higher-order cross-polarized $\mathrm{LG}_{nl}$ modes can produce a tangential energy flux in the same plane ($\phi = \pi/2$ or $3\pi/2$) where the HFGW signal can be optimally selected by the ideal GB in the Li-Baker detector~\cite{Li2003PRD}.

Under the condition $z=0$ and $\phi=\pi/2$, we have
\begin{align}\label{Sxy0pi2}
    \left\langle S_{\phi,(n,l)}^{xy}\right\rangle&\bigg|_{z=0,\phi=\pi/2}
=
\frac{r}{\mu_0\omega_e W_0^2}
A_{00}(r,0)A_{nl}(r,0)\notag\\
&\times\sin\left(\theta_{nl}-\theta_{00}-\frac{l\pi}{2}\right).
\end{align}
Using Eqs.\eqref{psi0} and~\eqref{eq:sigma}, Eq.~\eqref{Sxy0pi2} can be further simplified to
\begin{align}\label{Sxy0pi2expand}
    &\left\langle S_{\phi,(n,l)}^{xy}\right\rangle\bigg|_{z=0,\phi=\pi/2}=
\frac{r \psi_0^2}{\mu_0\omega_e W_0^2}
\exp\left[-\frac{2r^2}{W_0^2}\right]
\notag\\
&\times\sqrt{\sigma_{nl}}\,\mathcal M_{nl}\!\left(\frac{r}{W_0}\right)
\sin\left(\theta_{nl}-\theta_{00}-\frac{l\pi}{2}\right),
\end{align}
where
\begin{align}
     & \mathcal  M_{nl}\left(\frac{r}{W_0}\right)=\sqrt{\frac{n!}{(n+|l|)!}}
\left(\frac{\sqrt{2}\, r}{W_0}\right)^{|l|}\notag\\
&\times L_n^{|l|}\!\left[\left(\frac{\sqrt{2}\,r}{W_0}\right)^2\right].
\end{align}

Comparing Eq.~\eqref{Sxy0pi2expand} with Eq.~\eqref{S_ideal}, one can conclude that if  higher-order cross-polarized  $\mathrm{LG}_{nl}$ modes are present in the Li-Baker detector, the corresponding noise-to-signal ratio at $z=0$ and $\phi=\pi/2$ satisfies
\begin{align}\label{NSR41}
&\mathcal R_{nl}\left(\frac{r}{W_0}\right)
\equiv
\frac{\left|\left\langle S_{\phi,nl}^{xy}\right\rangle\right|}
{\left|\left\langle S_{\phi,\mathrm{GW}}^{(1)}\right\rangle\right|} =
\frac{4\psi_0}{A_{\times}\hat B_y^{(0)}\ell\,\omega_e}
\exp\left[-\frac{r^2}{W_0^2}\right]\,\notag\\
& \times
\sqrt{\sigma_{nl}}\left|
\mathcal M_{nl}\left(\frac{r}{W_0}\right)
\sin\left(\theta_{nl}-\theta_{00}-\frac{l\pi}{2}\right)
\right|.
\end{align}

It should be noted that a practical detector measures the integrated photon flux over a finite receiving area of the signal receiver. Consequently, the noise-to-signal ratio is typically defined as the ratio of the total noise photon flux to the signal photon flux across the receiving area. When the area is sufficiently small, the local noise-to-signal ratio can be reasonably approximated by the point-wise one. Using Eq.~\eqref{NSR41}, we further numerically assess the influence of higher-order cross-polarized LG beams on the noise-to-signal ratio.

\section{Numerical simulations}\label{sec4}

In this section, we conduct numerical simulations based on the energy-flux noise model presented in the previous section. Since the noise-to-signal ratio for higher-order co-polarized LG beams identically vanishes in the plane $\phi = \pi/2$, we primarily focus on higher-order cross-polarized LG beams here. We first analyze the contribution of an individual cross-polarized  $\mathrm{LG}_{nl}$ mode to the energy-flux noise, and then outline the limitations imposed by higher-order cross-polarized LG beams from Monte Carlo simulations. Finally, we discuss how to suppress the noise.

\subsection{Contribution of an individual  $\mathrm{LG}_{nl}$ mode}\label{subsec41}
To quantitatively calculate the noise-to-signal ratio for higher-order cross-polarized LG beams, we must specify the parameter values in Eq.~\eqref{NSR41}. The parameters used for numerical simulations are listed in Table~\ref{tab:params}. Here, the HFGW frequency ($\nu=\omega_e/2\pi=3.0\times10^{10}\,\mathrm{Hz}$) is chosen to be in the microwave band, and the strain amplitude (\(A_\times=10^{-30}\)) is a representative value widely adopted in HFGW feasibility studies.

\begin{table}
\caption{Parameters used for numerical simulations}
\centering
\begin{tabular}{l c c}
\hline
Quantity & Symbol & Value \\
\hline
HFGW frequency & \(\nu\) & \(3.0\times10^{10}\,\mathrm{Hz}\) \\
Angular frequency & \(\omega_e\) & \(2\pi\nu\) \\
Beam waist radius at $z=0$ & \(W_0\) & \(0.05\,\mathrm{m}\) \\
Static magnetic field & \(\hat B_y^{(0)}\) & \(30\,\mathrm{T}\) \\
Interaction length & \(\ell\) & \(0.10\,\mathrm{m}\) \\
HFGW strain amplitude & \(A_\times\) & \(10^{-30}\) \\
Peak amplitude of GB & \(\psi_0\) & \(3.0\times10^5\,\mathrm{V/m}\) \\
Evaluation radius & \(r\) & \(W_0/\sqrt{2}\) \\
\hline
\end{tabular}
\label{tab:params}
\end{table}

\begin{figure*}[t]
    \centering
    \includegraphics[width=0.9\textwidth]{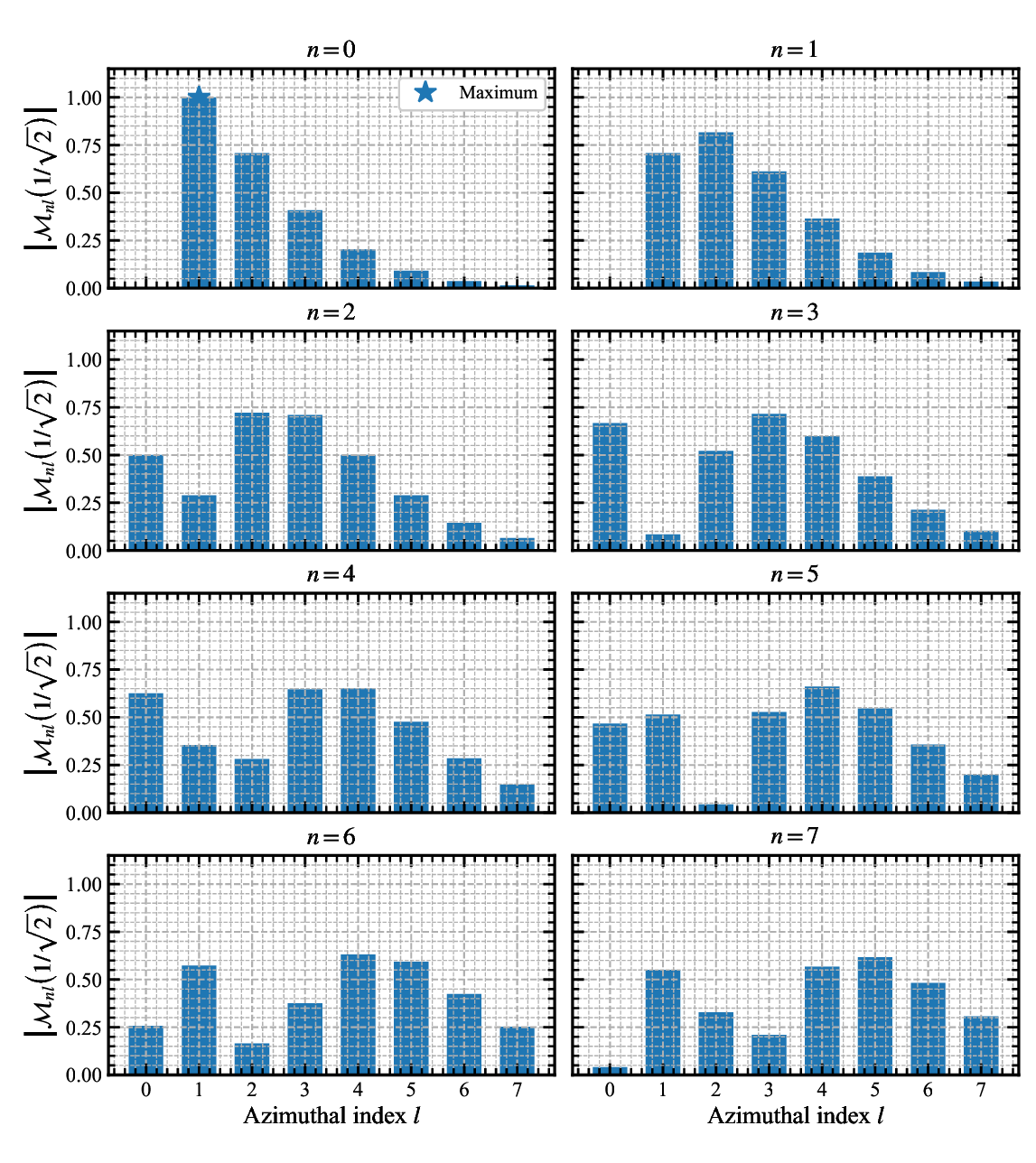}
    \caption{Values of the dimensionless factor $\left|\mathcal M_{nl}\left(\frac{1}{\sqrt{2}}\right)\right|$  for different higher-order cross-polarized  $\mathrm{LG}_{nl}$ modes at \(r=W_0/\sqrt{2}\).}\label{fig:M_nl}
\end{figure*}

Using the parameters listed in Table~\ref{tab:params}, the prefactor in Eq.~\eqref{NSR41} evaluates to
\begin{equation}
\frac{4\psi_0}{A_{\times}\hat B_y^{(0)}\ell\,\omega_e}\exp\left(-\frac{1}{2}\right)
=1.287\times10^{24}.
\label{eq:prefactor}
\end{equation}
Therefore, the noise-to-signal ratio for an arbitrary higher-order cross-polarized  $\mathrm{LG}_{nl}$ mode at \(r=W_0/\sqrt{2}\) can be expressed as
\begin{align}\label{upR43}
\mathcal R_{nl}\left(\frac{1}{\sqrt{2}}\right)
&=1.287\times10^{24}
\sqrt{\sigma_{nl}}\,
\left|
\mathcal M_{nl}\left(\frac{1}{\sqrt{2}}\right)
\right|,
\end{align}
where we have assumed $\sin\left(\theta_{nl}-\theta_{00}-\frac{l\pi}{2}\right)=1$. For an individual  $\mathrm{LG}_{nl}$ mode, the dimensionless factor $\left|\mathcal M_{nl}\left(\frac{1}{\sqrt{2}}\right)\right|$ can be numerically evaluated, with representative results shown in Fig.~\ref{fig:M_nl}. In this case, for arbitrary $(n,l)$, the corresponding noise-to-signal ratio
can be evaluated provided that the value of
{$\sigma_{nl}$}
is known. Conversely, the power fraction $\sigma_{nl}$ can be constrained by applying the noise-to-signal ratio criterion  $\mathcal R_{nl}<1$. Note that Eq.~\eqref{upR43} essentially represents the upper bound on the
noise-to-signal ratio for arbitrary $(n,l)$  and the phase term $\sin\left(\theta_{nl}-\theta_{00}-\frac{l\pi}{2}\right)$, thereby yielding a more stringent constraint
on {$\sigma_{nl}$} under the criterion $\mathcal R_{nl}<1$.

Figure~\ref{fig:M_nl} shows that the values of $\left|\mathcal M_{nl}\left(\frac{1}{\sqrt{2}}\right)\right|$
are mostly on the order of $10^{-1}$, which means that
\begin{align}
    \mathcal R_{nl}/\sqrt{\sigma_{nl}} \approx 10^{23} .
\end{align}
The criterion $\mathcal R_{nl}<1$ imposes a stringent upper bound on the power fraction of most individual $\mathrm{LG}_{nl}$ modes, requiring $\sigma_{nl}<10^{-46}$. Moreover, the largest individual contribution arises from the \(\mathrm{LG}_{01}\) mode,
for which $\left|\mathcal M_{01}\left(\frac{1}{\sqrt{2}}\right)\right|$ is exactly unity. Therefore, the individual $\mathrm{LG}_{01}$ mode yields the most stringent upper bound on the power fraction, with $\sigma_{01}<6.0\times 10^{-49}$. This implies that, for an individual higher-order cross-polarized $\mathrm{LG}_{nl}$ mode, its power contribution must be constrained to below approximately $1/10^{49}$ of the power of the fundamental Gaussian mode, so as not to overwhelm the signal in the Li-Baker detector.

\subsection{Monte Carlo simulations}\label{subsec42}

In this subsection, we conduct Monte Carlo simulations to determine
the relationship between the noise-to-signal ratio and the total power fraction of multiple higher-order cross-polarized $\mathrm{LG}_{nl}$ modes, thereby establishing the purity requirements for the GB adopted in the practical experiment.

Here, we restrict our attention to higher-order cross-polarized  $\mathrm{LG}_{nl}$ modes with $n < 7$ and $l < 3$. For larger $n$ and $l$, their contributions and effects are typically smaller, and are thus omitted from this analysis. To ensure that $\sqrt{\sigma_{nl}}$ is strictly positive, we define another parameter $a_{nl}$, which satisfies
\begin{align}
    \left|a_{nl}\right|=\sqrt{\sigma_{nl}}.
\end{align}
We assume that $a_{nl}$ is drawn from a normal distribution, and the generated random variables are subsequently normalized to satisfy
\begin{align}
    \sum_{(n,l)\ne(0,0)}a_{nl}^2=\sigma_{\mathrm{total}}.
\end{align}
Here, $\sigma_{\mathrm{total}}$ denotes the total power fraction of these higher-order cross-polarized $\mathrm{LG}_{nl}$ modes. For a given $\sigma_{\mathrm{total}}$, the distribution of $\left|a_{nl}\right|$ (and hence that of $\sqrt{\sigma_{nl}}$) is uniquely determined. In our numerical simulations, the value of $\sigma_{\mathrm{total}}$ ranges from $10^{-49}$ to $10^{-1}$ in increments of one order of magnitude. Since the energy-flux noise contributions from individual $\mathrm{LG}_{nl}$ modes are statistically independent, the total noise-to-signal ratio (denoted as $\mathcal R_\mathrm{total}$) can be evaluated using the standard error propagation formula, yielding
\begin{align}
    \mathcal R_\mathrm{total}=\sqrt{\sum_{(n,l)\ne(0,0)}\mathcal R_{nl}^2}.
\end{align}

\begin{figure}[h]
    \centering
    \includegraphics[width=0.5\textwidth]{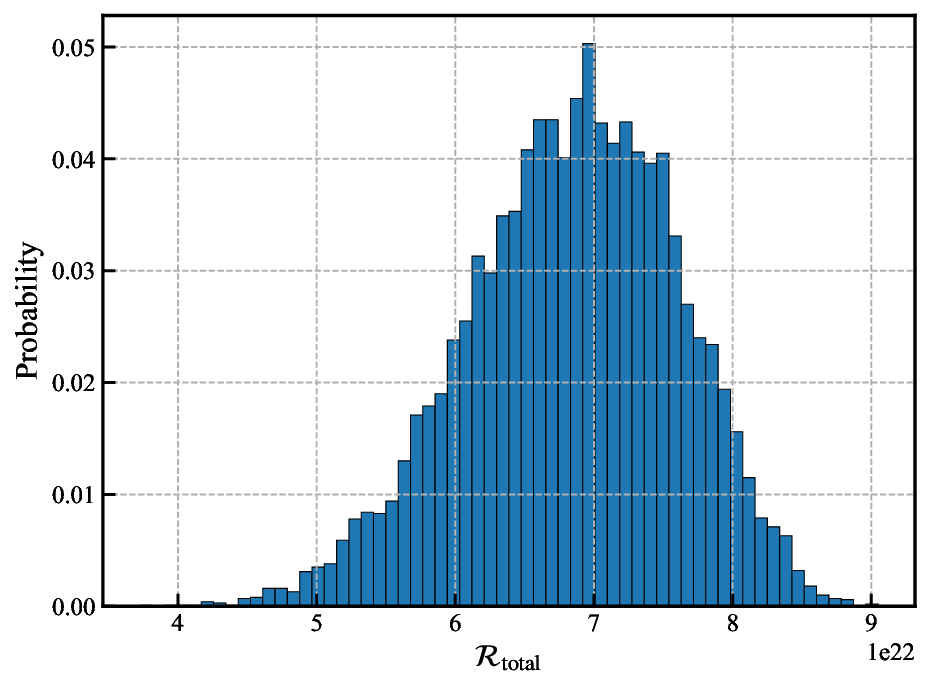}
    \caption{Probability distribution of the total noise-to-signal ratio from Monte Carlo simulations. With the total power fraction of multiple higher-order cross-polarized $\mathrm{LG}_{nl}$ modes set to $\sigma_{\mathrm{total}}=10^{-3}$, the distribution is obtained by randomly assigning power fractions to individual $\mathrm{LG}_{nl}$ modes.}
    \label{fig:1e-3}
\end{figure}

\begin{figure}
    \centering
    \includegraphics[width=0.5\textwidth]{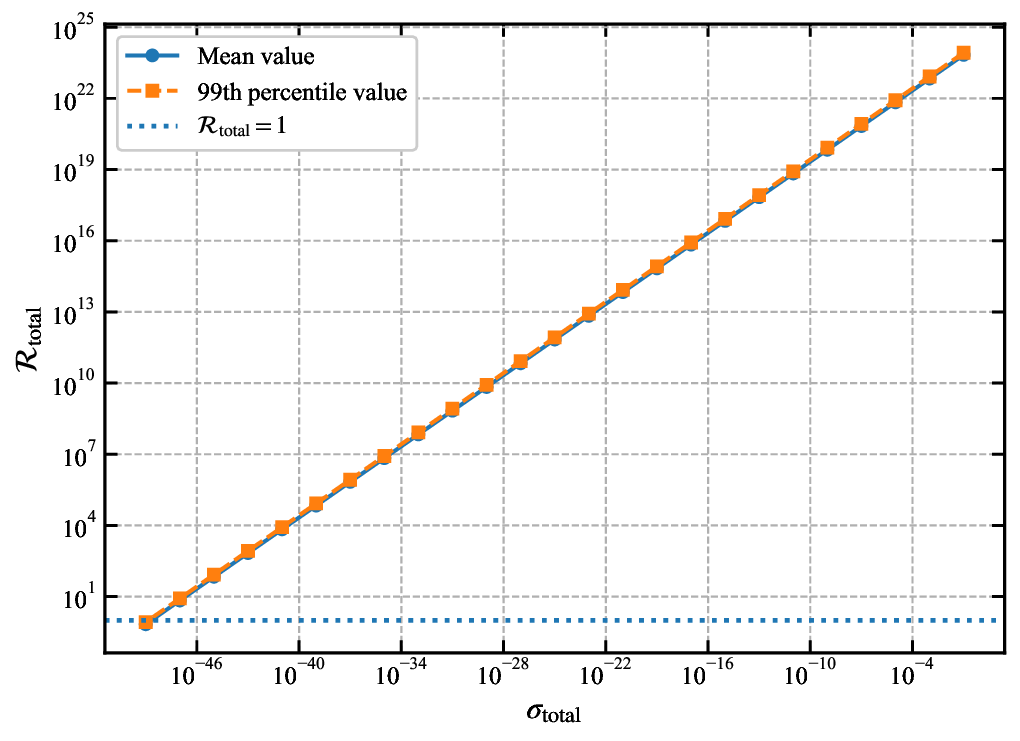}
    \caption{Log-log plot of the noise-to-signal ratio versus $\sigma_\mathrm{total}$.}
    \label{fig:99mean}
\end{figure}

In Fig.~\ref{fig:1e-3}, we present the distribution of $\mathcal R_\mathrm{total}$ for a total power fraction of $\sigma_{\mathrm{total}}=10^{-3}$. Notably, this distribution is representative of other values of $\sigma_{\mathrm{total}}$ as well.
The shape of the distribution remains invariant as the total power fraction varies,
with only the scale of the horizontal axis changing accordingly.
Note that the resulting magnitude of $\mathcal R_\mathrm{total}$
is on the order of $10^{22}$, which far exceeds the requirement that the
noise-to-signal ratio should be smaller than unity. Figure~\ref{fig:99mean}
presents the mean and 99th-percentile values \(\mathcal R_\mathrm{total}\) obtained
from 10000 random realizations, with the horizontal dashed line indicating the threshold \(\mathcal R_\mathrm{total}=1\). For the adopted parameters, the results indicate that the detection of HFGWs can remain unaffected only when  the total power fraction of these higher-order cross-polarized $\mathrm{LG}_{nl}$ modes satisfies the condition $\sigma_{\mathrm{total}} <10^{-49}$. This result is consistent with the conclusions on an individual $\mathrm{LG}_{nl}$ mode presented in the preceding subsection.

The numerical simulations demonstrate that higher-order cross-polarized LG beams have a profound impact on the noise-to-signal ratio of the Li-Baker detector. When these modes are present, the purity of the fundamental Gaussian mode would be required to reach an extreme level of $1-10^{-49}$. Since such a stringent requirement is practically unattainable,  eliminating higher-order cross-polarized LG beams is a critical consideration for the Li-Baker detector.

\section{Conclusion and Discussion}\label{sec5}
In this work, we have systematically investigated the energy-flux noise induced by imperfections of the GB in the Li-Baker detector. Using higher-order LG beams to model the GB imperfections, we analytically derived the resulting energy-flux noise and evaluated its impact on the noise-to-signal ratio.

Our findings reveal a stark asymmetry between polarization states of higher-order LG beams.
Higher-order co-polarized $\mathrm{LG}_{nl}$ modes ideally do not generate energy-flux noise
in the optimal detection plane ($\phi = \pi/2$ or $3\pi/2$), whereas cross-polarized $\mathrm{LG}_{nl}$ modes introduce significant noise. Through Monte Carlo simulations, we demonstrated that if multiple higher-order cross-polarized $\mathrm{LG}_{nl}$ modes are present in the Li-Baker detector, the total power fraction  $\sigma_{\text{total}}$ of these modes must be stringently constrained to below $10^{-49}$ to ensure that the noise-to-signal ratio remains below unity. This is a very stringent constraint compared to the approximately 0.5\% higher-order-mode power fraction reported for LIGO.

To achieve successful HFGW detection in the Li-Baker detector, it is essential to implement control measures to reduce the total power fraction of higher-order cross-polarized LG beams. An effective strategy is to pass the imperfect GB through an ideal polarizer before it enters the region containing the background static magnetic field, thereby filtering out all higher-order cross-polarized LG beams and retaining the co-polarized component. This ideal polarizer neither perturbs the existing background static magnetic field nor interferes with the incident HFGWs. Crucially, since the fundamental Gaussian mode is also co-polarized, the polarizer leaves signal generation unaffected. Although such an ideal polarizer may not exist, this strategy is still expected to help suppress the energy-flux noise induced by the GB imperfections and thereby facilitate HFGW detection in the Li–Baker detector.

\bmhead{Acknowledgements}
This work is supported by the National Key R$\&$D Program of China under Grant No. 2023YFC2205500, and the National Natural Science Foundation of China under Grants No. 12405060 and No. 12505058.

\bibliography{reference}

\end{document}